\documentclass[a4paper,11pt]{article}
\usepackage{jinstpub} 

\usepackage{graphicx}%
\usepackage{amsmath}
\usepackage{amsthm}%
\usepackage{wasysym} 
\usepackage{xcolor}%

\usepackage{lineno}

\title{\boldmath Detecting photons in radiative Bhabha interaction with Si-strip  and  LYSO crystals at future $e^+e^-$ colliders}

\author[a]{Chentao~Bao,}
\author[a]{Hai~Chen,}
\author[b]{Wang~Chen,}
\author[c]{Zhujun~Fang,}
\author[d]{Jiading~Gong,}
\author[e,1]{Suen~Hou,\note{Corresponding author.}}
\author[f]{Xiaolin~Kang,}
\author[c]{Mingyi~Liu,}
\author[g]{Ming~Qi,}
\author[h]{Liqing~Qin,}
\author[d]{Weimin~Song,}
\author[f]{Jie~Wang,}
\author[b]{Yadi~Wang,}
\author[g]{Lei~Zhang,}
\author[h]{Linghua~Zhang}

\affiliation[a]{Zhejiang University, Hangzhou 310058, China}
\affiliation[b]{North China Electric Power University, Beijing 102206, China}
\affiliation[c]{University of Science and Technology of China, Hefei 230026, China}
\affiliation[d]{Jilin University, Changchun 130012, China}
\affiliation[e]{Academia Sinica, Taipei 115201, China}
\affiliation[f]{China University of Geosciences, Wuhan 430074, China}
\affiliation[g]{Nanjing University, Nanjing 210093, China}
\affiliation[h]{Guangxi Normal University, Guilin 541004, China}

\emailAdd{suen@as.edu.tw}

\abstract{
Next-generation $e^+ e^-$ colliders have been proposed 
for the Super Tau Charm Facility covering center-of-mass energies
 $\sqrt{s} =2 - 7$~GeV, and for the Higgs Factories 
optimized for Z-pole physics and for HZ production.
Luminosity can be determined by measuring forward Bhabha scattering.
The QED cross-section calculation has reached a precision of better than $10^{-3}$.
Detecting radiative photons in Bhabha events
provides a tool for validating higher-order corrections in theory.
We propose the Forward Detector for detecting Bhabha scattering events at the STCF.
The design uses high-resolution silicon detectors and finely segmented 
LYSO crystals to discriminate radiative photons from closely associated electrons.
Bhabha scattering events are simulated with the BHLUMI and BHWIDE programs,
and the detector is modeled using GEANT. 
Systematic uncertainties due to multiple scattering and bremsstrahlung 
in the beam pipe were evaluated.
The potential for achieving precise measurements of 
radiative processes in Bhabha events is investigated.
}


\begin{document}
\maketitle
\flushbottom

\section{Introduction}
\label{sec:intro}

Luminosity measurements at future $e^+ e^-$ colliders are proposed for detecting
Bhabha elastic scattering events in forward polar-angle regions  
for the Flavor Facility at the
Super Tau Charm Facility (STCF) \cite{STCF-CDF} and the Higgs factories 
CEPC  \cite{CEPC-TDR} and FCC-ee \cite{FCC-ee}. 
The physics programs 
require luminosity measurements with a precision better than $10^{-3}$.
Identification of $e^\pm /\gamma$ in radiative $e^+ e^- \to e^+ e^- (n\gamma)$ 
interaction will measure higher-order processes predicted by QED.
The Bhabha event generators BHLUMI \cite{BHLUMI404} 
and BHWIDE \cite{BHWIDE100} are employed to study
acceptances in the forward region of both STCF and Higgs factories.

\begin{figure}[b!]  
  \vspace{-2mm}
  \centering  
  \includegraphics[width=.65\linewidth]{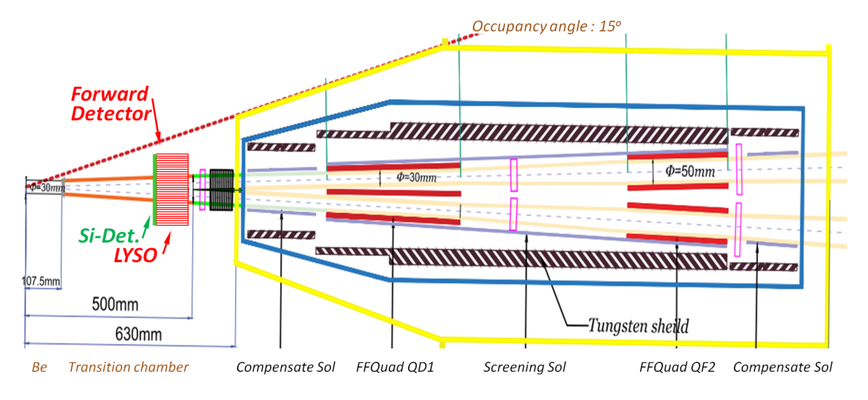} 
  \vspace{-5mm}
  \caption{  \label{fig:STCF_MDI}
  The FWD on each z-side of the Machine Detector Interface (MDI) 
  region is shown. The FWD consists of silicon sensors and 
  LYSO arrays mounted before the beam-pipe flanges.
  }
\end{figure}  

The detection of Bhabha events is investigated using the Forward Detector (FWD)
design shown in Fig.~\ref{fig:STCF_MDI}.
It is located in the Machine Detector Interface (MDI) region to
measure $e^+ e^-$ collisions over center-of-mass energies  
of $\sqrt{s}=2 -7$~GeV with a 60 mrad beam-crossing angle.
The beam pipe is a circular Be cylinder at the interaction point (IP) 
with an inner radius of $r=15$~mm and a thickness of 2 mm.
It expands horizontally into a racetrack shape 
toward the MDI flanges, then splits into two distinct 
round pipes before reaching the quadrupole magnets.
The laboratory frame has the $z$-axis aligned between the beam pipes,
the $y$-axis points upward, and the beam crossing boosts toward the positive $x$-direction.
The detector components consist of Si-strip 
detectors and finely segmented LYSO  
for exclusive measurement of $e^\pm$ and $\gamma$ in radiative Bhabha scattering 
events.  
The $e^\pm$ and $\gamma$ shower profiles are simulated using 
GEANT~\cite{GEANT3, GEANT4}.
The systematics arising from multiple scattering and bremsstrahlung 
in the beam pipe are also investigated.

\begin{figure}[b!]  
  \vspace{-2mm}
  \centering  %
  \includegraphics[width=.75\linewidth]{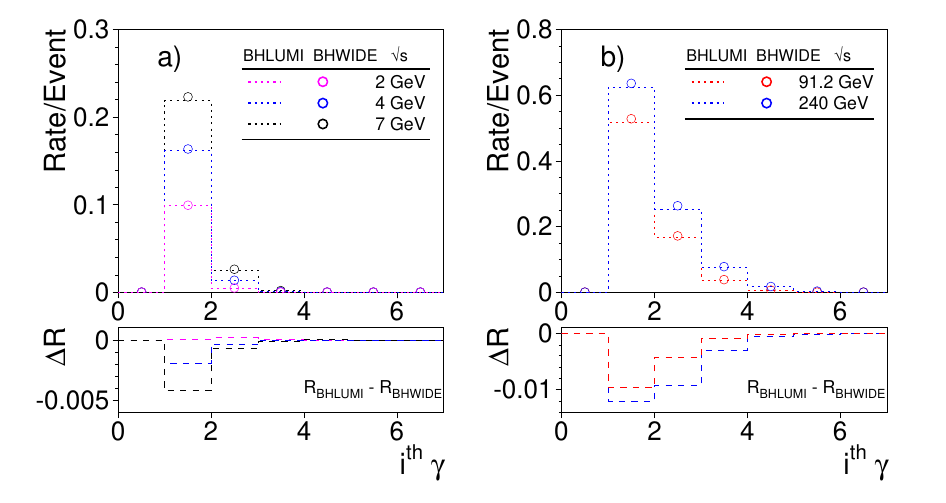} 
  \vspace{-6mm}
  \caption{  \label{fig:Ngamma}
   Radiative Bhabha events with $e^+$ and $e^-$ in the $\theta$
   range of $1^\circ - 20^\circ$ are 
   plotted for event rates, with photons of $E_\gamma > 0.1$ GeV 
   counted in descending order of $E_\gamma$, for $\sqrt{s}$ 
   at STCF (2 -7 GeV) and at 
   Higgs factories. The differences $(\Delta R)$ between 
   BHWIDE and BHLUMI are also plotted.
  }

  \vspace{2mm}
  \centering  %
  \includegraphics[width=.75\linewidth]{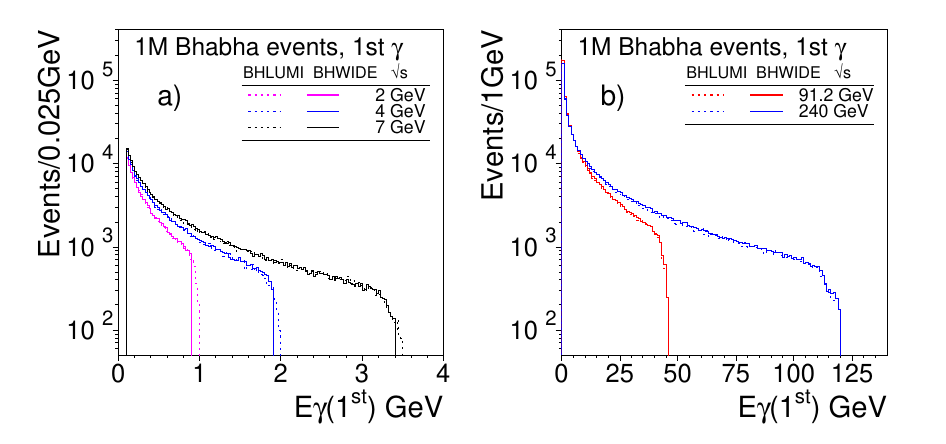} 
  
  \vspace{-5mm}
  \caption{  \label{fig:Egamma}   
   The energy distributions of the most energetic photon ($1^{\rm st} \gamma$, 
   $E_\gamma >0.1$~GeV) are plotted. 
   The statistics correspond to 1M Bhabha events generated  
   in $1^\circ - 20^\circ$. The event rates at $E_\gamma$ near zero 
   are compatible for different $\sqrt{s}$ and extend to $E(e^\pm)$.
  }
\end{figure}  

\section{Acceptance of radiative Bhabha event in the FWD}
\label{sec:Accept}

Radiative Bhabha interactions are simulated with the
BHLUMI and BHWIDE programs, 
which employ the Yennie-Frautschi-Suura (YFS) formalism 
for soft-photon resummation \cite{YFS}. 
Events are generated for both $e^+$ and $e^-$ restricted to the
forward $\theta$ range of $1^\circ$ to $20^\circ$ 
in the center-of-mass (CM) frame at selected energies.
The programs have their default physics parameters employed (as in demo.f). 
The generation of radiative photons in $e^+ e^- \to e^+ e^- (n\gamma)$ 
has energy thresholds of $10^{-4}$ ($10^{-5}$) of $E(e^\pm)$ in 
BHLUMI (BHWIDE), respectively. 

The multiple particle final states of $e^+ e^- (n\gamma)$
cause deviation of $E(e^\pm)$ from $e^\pm$ beam energy $(\sqrt{s}/2)$
and acollinearity angle of $e^+$ and $e^-$.
Photons are classified as initial- and final-state
radiation (ISR, FSR) according to the opening angle $\Omega(e^\pm,\gamma)$
being closer to the incident $e^\pm$ (beam directions)
or the scattered $e^\pm$. 
Photons in each event are counted in descending order of energy 
with a threshold of $E_{\gamma}>0.1$~GeV.
In Fig.~\ref{fig:Ngamma}, the event rates of photons counted in 
sequence are plotted for the $\sqrt{s}$ energies at both STCF and Higgs factories.

The energy spectra of the first radiative photons are plotted in 
Fig.~\ref{fig:Egamma}. 
The distributions are generated with 1M Bhabha events 
(corresponding to the event rate in the $N=1$ bin of Fig.~\ref{fig:Ngamma}).
The photon event rates for $E_\gamma$ close to zero are compatible with
different $\sqrt{s}$ values.  
As $\sqrt{s}$ increases, the event rate with radiative photons rises 
by allowing photon emission at energies up to $E(e^\pm)$.

\begin{figure}[b!]  
  \centering  %
  \hspace{3mm}\includegraphics[width=.415\linewidth]{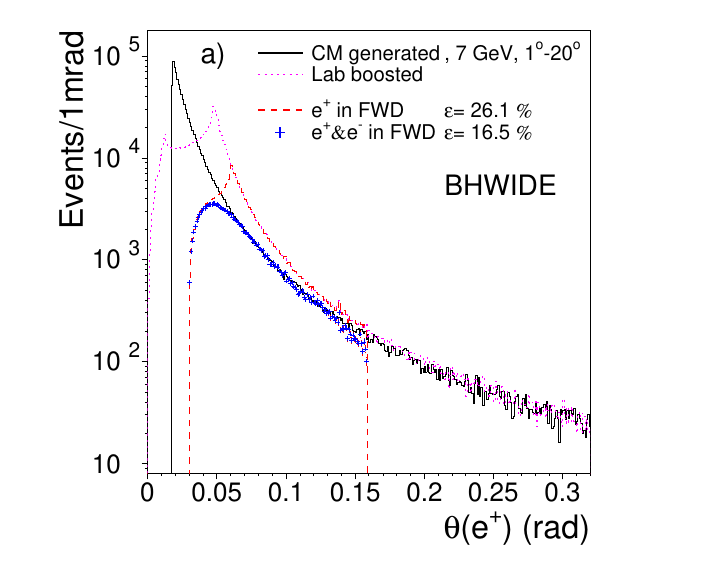}\hspace{-10mm}\includegraphics[width=.415\linewidth]{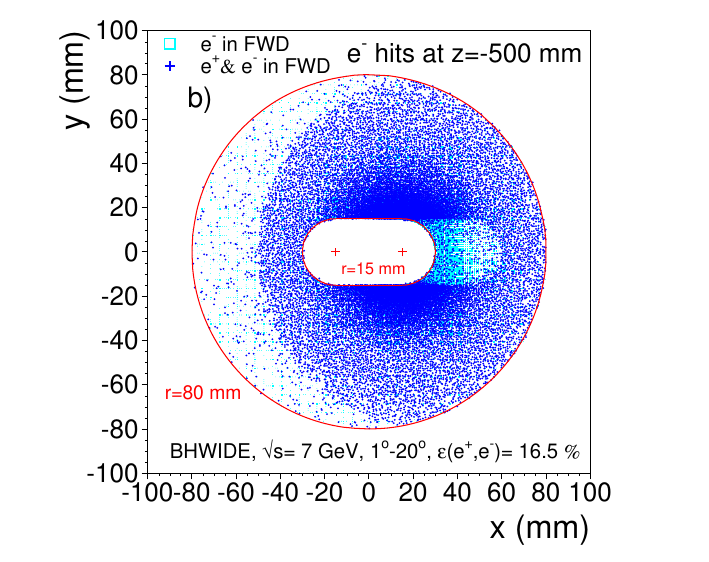}
  
  \vspace{-4mm}
  \caption{  \label{fig:thetaxy}
  The BHWIDE generated Bhabha events at $\sqrt{s} = 7$~GeV in the $\theta$ range of 
   $1^\circ - 20^\circ$ (1M events) are plotted for
   a) $\theta$ distributions of scattered $e^+$
   in the CM frame, and the boosted in laboratory
   (by the beam-crossing, magenta dotted) selected in the FWD acceptance 
   with single $e^+$ or both $e^+$ and $e^-$ 
   detected, respectively;
   b) the plenary distribution of $e^-$ hits 
   with $e^-$ only (cyan) or both $e^+$, $e^-$  (blue) 
   detected in the FWD acceptances (red lines) are plotted for the projection at 
   $|z| = 500$ mm. 
}
\end{figure}  

The Bhabha cross section has a $1/s$ dependence 
and a steep $1/\theta^2$ distribution 
of the scattered $e^\pm$. 
The $e^+ e^-$ collision luminosity is measured with the counted 
Bhabha events ($N_{\rm acc}$) versus the Bhabha cross section ($\sigma$). 
The integrated luminosity $\mathcal{L}$ is given by:   
\begin{eqnarray}
  \mathcal{L} &=& \frac{1 }{ \epsilon }  
          \frac{ N_{acc}}{\sigma } \label{intL} \\
  \sigma &=& \frac{16 \pi \alpha^2}{ s} \left(
          \frac{1}{\theta_{min}^2} - \frac{1}{\theta_{max}^2} \right). \label{BhaXsec}  
\label{eq:BabaSigma}          
\label{eq:dLL} 
\end{eqnarray}
Fig.~\ref{fig:thetaxy}a illustrates 
the $\theta$ distributions of scattered electrons generated 
by the BHWIDE at $\sqrt{s}=7$~GeV in the CM frame, 
which are then boosted to the laboratory frame by the beam crossing.
Bhabha event acceptance is most sensitive to $\theta_{\rm min}$ 
where a shift in $\theta_{\min}$ leads to a relative luminosity error of
$\Delta \mathcal{L}/ \mathcal{L} = 2\, \Delta\theta/\theta_{\rm min}$.
For the luminosity measurement precision of $10^{-4}$, the corresponding
deviation would be $1.5 \; \mu$rad at $\theta_{\rm min}=30$~mrad.

The Forward Detector acceptance is investigated for the 
region of $r< 80$~mm at $|z|=500$~mm from the IP, 
outside the beam pipe, with centers at $x=\pm 15$~mm and $r=15$~mm.
In Fig.~\ref{fig:thetaxy}a, the boosted $e^+$ distributions
are also plotted for $e^+$ in the FWD region,
and for both $e^+$ and $e^-$ selected, respectively.
The distributions plotted in the $x$-$y$ plane are shown in 
Fig.~\ref{fig:thetaxy}b. 

The cross sections listed in Table~\ref{tab:Acc_FWD} are calculated 
for $e^+$ and $e^-$   
polar angles in the range of $1^\circ$–$20^\circ$ in the CM frame.
In this angular range, BHLUMI and BHWIDE agree within 0.1~\% 
across the center-of-mass energies of STCF and Higgs factories.
The acceptance in the FWD, with $\theta_\pm^{\rm bp} >30$~mrad 
relative to the beam centers
at $|z|=500$~mm within $\theta_{\rm lab} <160$~mrad, are also listed.
The cross sections of BHWIDE are slightly higher by about 2~\% 
at the STCF energies, and 5~\% at Higgs factories.

Events containing radiative photons are selected with $E_\gamma >0.1$~GeV 
within the FWD regions on either side of the $z$-hemispheres. 
The event rates ($R_{\pm z}(\gamma^{1\mathrm{st}})$) are also listed in Table~\ref{tab:Acc_FWD}.
For the ISR photons, the event rate is around 0.5~\% at the STCF 
and increases to 1~\% at Higgs factories. 
FSR photons are emitted in close alignment with the scattered electrons.
Assuming that the segmentation of calorimeters can distinguish 
$e^\pm/\gamma$ with an opening angle of $\Omega(e^\pm,\gamma)>10$~mrad,
the event rates to be detected reach 1.5~\% at STCF 
and rise to 3~\% at Higgs factories.

\begin{table*}[t!]
\vspace{2mm} 
\centering
{ 
\small

\begin{tabular}{lccc|cc}                   
\hline
\hline  
BHWIDE $\sqrt{s}$ (GeV)  &  2 & 4 & 7   & 91.2   & 240  \\
\hline
$\sigma(1^\circ$-$20^\circ)$ nb &
    854000 & 214600 & 70160         & 423.4     & 62.08  \\   
$\sigma$(FWD) nb &
    139900 & 35230  & 11570         & 71.93     & 10.65   \\
\hline   
$R_{\pm z}(\gamma^{1\mathrm{st}} \mathrm{ISR})$ & 
     .0040 & .0049  & .0060         & .0095     & .0103   \\
$R_{\pm z}(\gamma^{1\mathrm{st}} \mathrm{FSR},\Omega>\!{\rm10\,mrad})$ & 
     .0094 & .0126  & .0156         & .0262     & .0291  \\
\hline
\\
\hline 
BHLUMI $\sqrt{s}$ (GeV)  &  2 & 4 & 7   & 91.2   & 240   \\
\hline
$\sigma(1^\circ$-$20^\circ)$ nb &
    857800 & 214300 & 70260         & 424.0     & 62.08   \\ 
$\sigma$(FWD) nb & 
    137900 & 34640  & 11380         & 69.00     & 10.16  \\
\hline  
$R_{\pm z}(\gamma^{1\mathrm{st}} \mathrm{ISR})$ & 
     .0045 & .0054  & .0062         & .0096     & .0105  \\
$R_{\pm z}(\gamma^{1\mathrm{st}} \mathrm{FSR},\Omega>\!{\rm10\,mrad})$ & 
     .0092 & .0126  & .0153         & .0257     & .0283   \\
\hline
\hline
\end{tabular}
}

\vspace{2mm}
  \caption{    \label{tab:Acc_FWD} 
  Bhabha cross sections of BHWIDE and BHLUMI calculations are listed 
  for $\sqrt{s}$ from 2 to 240~GeV in $1^\circ - 20^\circ$ (CM frame), 
  and for the boosted case with $e^+$ and $e^-$ in FWD.
  The event rates containing radiative photons in the FWD are 
  also listed for events with the most energetic photon ($R_\pm(\gamma^{\rm 1st})$) 
  selected for $E_\gamma > 0.1$ GeV and the opening angle 
  $\Omega (e^\pm,\gamma) > 10$~mrad to the $e^\pm$ in the same $z$-hemisphere.  
  }
\end{table*}

\section{Forward Detector simulation and $e^\pm/\gamma$ separation }
\label{sec:GEANT}

\begin{figure}[b!]  
  \vspace{-2mm}
  \centering  %
  \includegraphics[width=.92\linewidth]{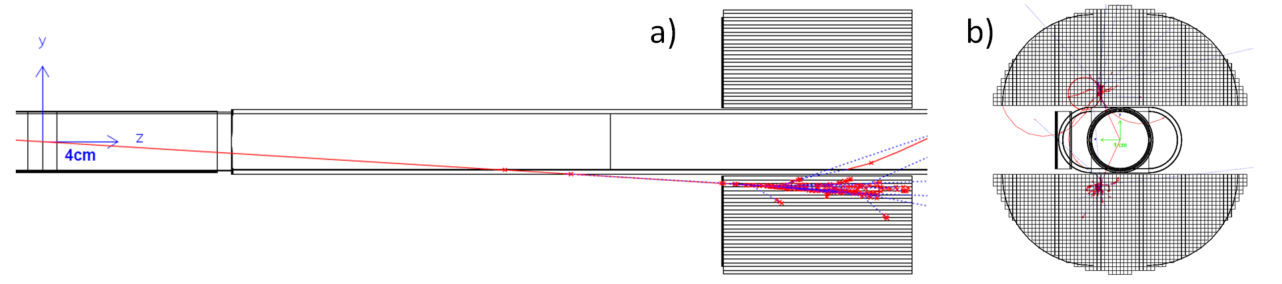} 
  \caption{  \label{fig:gin3}
  GEANT simulation is shown for a pair of 3.5 GeV $e^+$ and $e^-$
  injected vertically at 60 mrad to the $z$-axis from the IP, 
  passing through a 2 mm thick Be pipe to the FWD Si wafer before the LYSO at 
  $|z| = 360 - 460$ mm ($8.7 X_0$).  
}
\end{figure}

The Forward Detector design consists of a Si detector located 
in front of LYSO crystal arrays for measuring Bhabha scattered electrons 
and radiative photons. 
The LYSO crystal is positioned in the MDI region before the flanges 
at $|z|=460$ mm.  
A 100 mm LYSO length is used to capture a broader range of 
shower profiles in GEANT simulations.
The LYSO crystal is divided into $2\times 2\times 100$~mm$^3$ segments,
oriented parallel to the $z$-axis in the laboratory frame.
Fig.~\ref{fig:gin3} presents an event display in the FWD
for a pair of Bhabha-scattered $e^+$ and $e^-$ of 3.5~GeV.
The boost in the $+x$ direction is applied for the beam-crossing effect.

\begin{figure}[t!]  
  \centering  
  \hspace{4mm}\includegraphics[width=.8\linewidth]{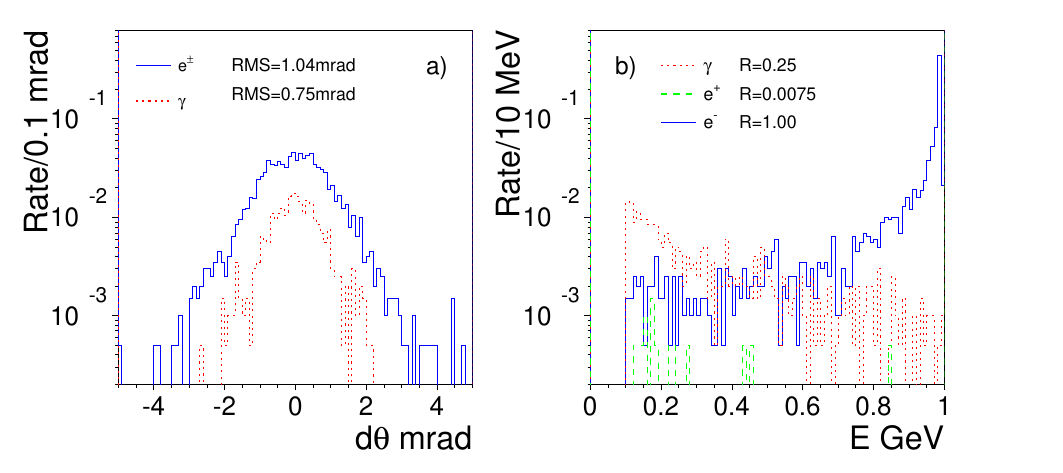}
  
  \vspace{-6mm}
  \caption{  \label{fig:msps}
  The GEANT simulations had 1 GeV electrons injected vertically at 
  $\theta=57$~mrad through a 2~mm Be pipe; the traversing distance corresponds to $0.1 X_0$.  
  The beam electrons and shower secondaries are 
  selected with $E(e^\pm), E_\gamma >0.1$~GeV.
  The distributions measured at the Si detector ($|z|=359$~mm) are normalized 
  to event rates.
  The deviations in $\theta$ caused by multiple scattering are plotted in a).
  The energy spectra are plotted in b). 
  The bremsstrahlung photons, occurring at a rate of R, are emitted along 
  with the beam electrons with an RMS of 0.7 mrad in $d\theta$.   
}
\end{figure}  

The beam pipe material upstream of the silicon detector and LYSO arrays 
degrades measurement precision by inducing multiple scattering 
and pre-shower bremsstrahlung.
GEANT simulations were performed for electrons traversing
a $\diameter 30$~mm, 2~mm thick Be pipe.
The distributions made with 1~GeV electrons are plotted 
in Figure~\ref{fig:msps} for the angular deflection 
($d \theta$) and energy spectra of the incident electrons 
and secondary particles.

The distributions of $e^\pm$ and $\gamma$ in 
Fig.~\ref{fig:msps} are selected for 0.1~GeV. 
The electron beam is injected at a vertical angle of 57~mrad; 
the path length corresponds to $0.1 \; X_0$.
The RMS of multiple scattering of the 1 GeV electrons 
is 1.04~mrad, and the bremsstrahlung photon rate is 25~\%.
With the GEANT default cut-offs of 1~MeV, 
the width in $d\theta$ matches that of the beam electrons,
and the number of photons increases fourfold.
As the beam energy increases, multiple scattering becomes narrower, 
and the bremsstrahlung rate increases.
Distributions made with 3.5 GeV electrons have 
an RMS of $d\theta$ equal to 0.40~mrad, 
and the bremsstrahlung rate of $E_\gamma >0.1$~GeV photons is 41~\%. 

Si strip or pixel sensors measure $e^\pm$ impact position 
with a spatial resolution of a few microns, 
whereas photons pass through without generating ionization signals.
Thus, placing Si detectors in front of the LYSO arrays
enables discrimination between $e^\pm$ and $\gamma$.
The dominant systematic uncertainties arise from secondary particles 
produced in the beam pipe, which remain tightly grouped within 
the multiple-scattering cone to form a single LYSO cluster.

\begin{figure}[t!]  
  \vspace{-4mm}
  \centering  
  \hspace{2mm}\includegraphics[width=.42\linewidth]{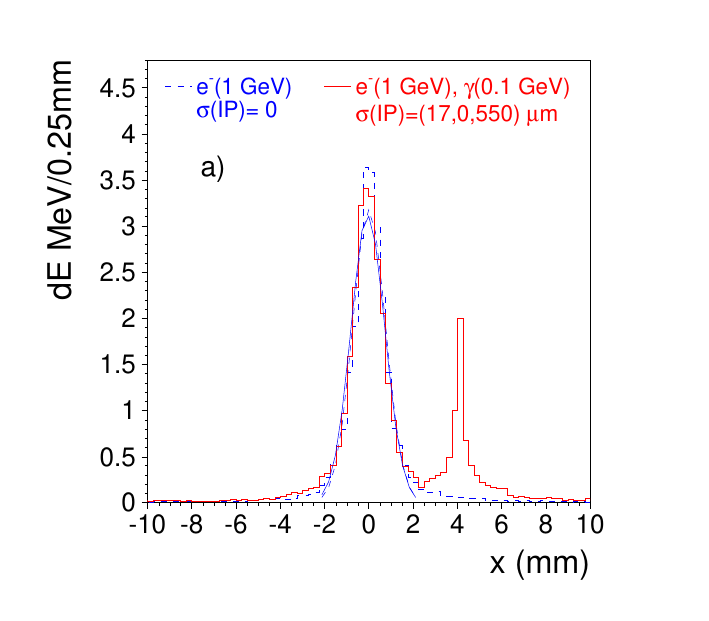}\hspace{-10mm}\includegraphics[width=.42\linewidth]{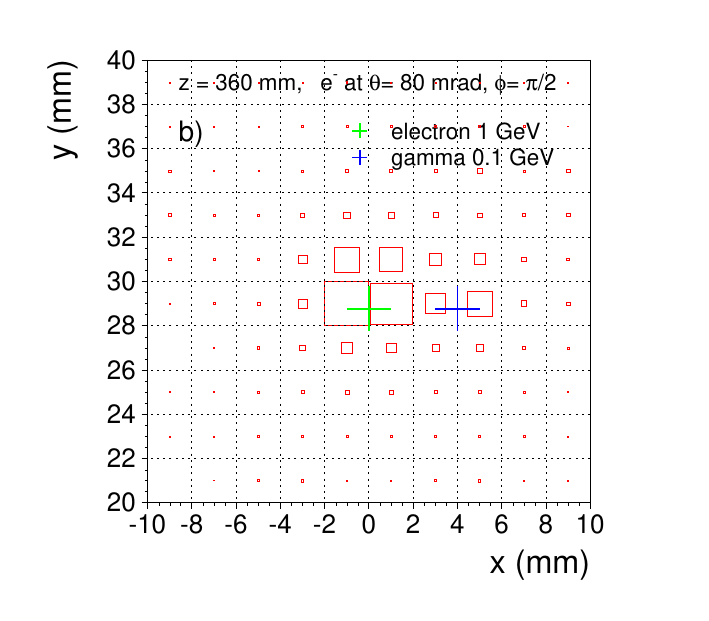}
  
  \vspace{-8mm}
  \caption{  \label{fig:eg_lat}
  The lateral shower profiles at the initial $1 \,X_0$ of LYSO are presented,
  normalized for a 1 GeV electron:
   a) with and without IP smearing for the colliding $e^\pm$ bunch size, 
   adding a 0.1 GeV photon separated by 4 mm; 
   b) deposits of 
  both $e^-$ and $\gamma$ accumulated in the $1 \, X_0$ $2\times2$ mm$^2$ 
  LYSO arrays; the impact positions are marked, which are between the LYSO cells.
}

  \centering  

  \hspace{2mm}\includegraphics[width=.42\linewidth]{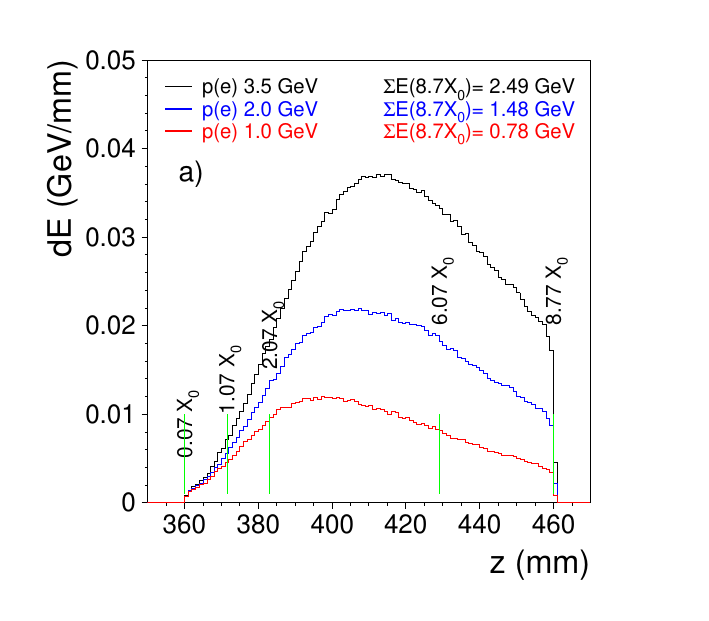}\hspace{-10mm}\includegraphics[width=.42\linewidth]{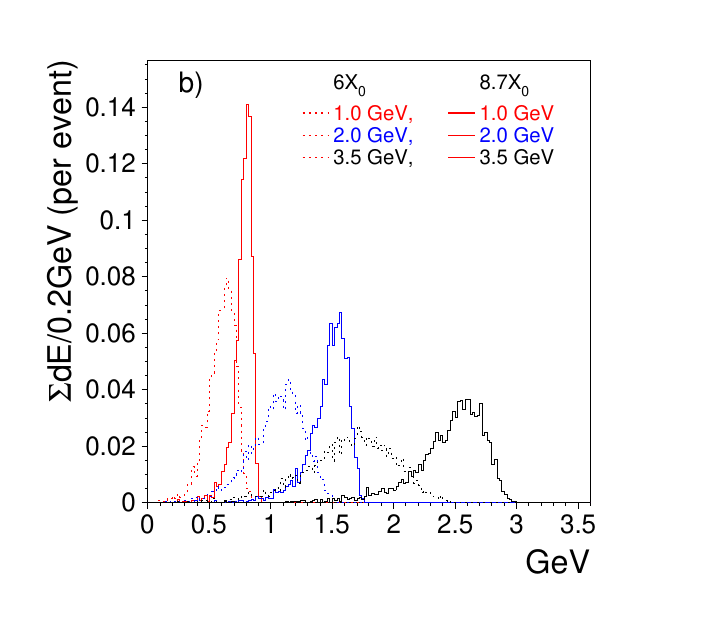}

  \vspace{-8mm}
  \caption{  \label{fig:eg_lyso}
  Electrons of 1, 2, and 3.5 GeV injected at 80 mrad vertically to the FWD, 
  after traversing a 2 mm thick Be pipe ($0.071 \, X_0$), are plotted for 
  a) longitudinal event profiles and b) $\Sigma dE$ distributions in 
  $6 \, X_0$ (dotted) and $8.7 X_0$ (solid) LYSO arrays.
}
\end{figure}

In a LYSO crystal, an electromagnetic shower expands conically 
with radiation length.
Sampling shower clusters in the first radiation length
improves the separation between adjacent $e^\pm$ and $\gamma$.
The simulations performed with the initial $1 \, X_0$ of LYSO are plotted 
for the lateral profiles in Fig.~\ref{fig:eg_lat}a.
Events were generated with 1 GeV electrons injected vertically 
at 80 mrad from the IP, accompanied by a
0.1 GeV photon offset by 4~mm from the $e^-$ at the front of the LYSO. 
The deposits of both $e^-$ and $\gamma$ accumulated in 
$2\times 2$ mm$^2$ arrays are shown in Fig.~\ref{fig:eg_lat}b.

The Bhabha scattered electrons are identified
with energy deposits in the calorimeter.
The 100~mm long LYSO arrays, located at $|z| = 360 -460$~mm, 
correspond to a radiation length of $8.7\,X_0$.
Fig.~\ref{fig:eg_lyso}a illustrates the longitudinal shower profiles 
along the LYSO length for 1, 2, and 3.5~GeV electrons
injected at 80~mrad vertically at $x=0$.
The 2~mm thick Be pipe up front corresponds to $0.07 \, X_0$. 
The shower maximum positions increase from  
$\sim 2.8\,  X_0$ to $\sim 4.4 \, X_0$
for electrons of 1 to 3.5~GeV, respectively.
The sums of energy deposits are plotted in Fig.~\ref{fig:eg_lyso}b.
The  mean (RMS) values are
0.77 (0.08),  1.47 (0.17), and 2.45 (0.31)~GeV, respectively. 
With a shorter LYSO length of $6 \, X_0$, 
the mean and RMS values become 67~\% and 117~\% of those at 
$8.7 X_0$ for 3.5~GeV electrons.

\section{ Summary }
\label{sec:sum}

The acceptance of forward-angle Bhabha scattering events 
is evaluated for the STCF and Higgs factories.
The detector configuration comprises
Si detectors and a finely segmented LYSO calorimeter.
The Si detectors track impact positions 
of charged $e^\pm$ particles, while a 1 $X_0$ sampling LYSO arrays 
distinguishes shower clusters between $\gamma$ and $e^\pm$.
Detecting radiative photons at the percent level may facilitate 
the validation of Bhabha scattering predictions to a precision 
of $10^{-3}$ or better.

\acknowledgments

The authors would like to thank the assistance
of the STCF accelerator group of the University of 
Science and Technology of China. 
This work has been partially supported by 
the Institute of Physics, Academia Sinica.




\begin{thebibliography}{}
 


\bibitem{STCF-CDF}
“STCF Conceptual Design Report: Volume 1 -- 
Physics \& Detector”, M. Achasov et al., Front Phys. 19, 14701 (2024)

\bibitem{CEPC-TDR}
“CEPC Technical Design Report -- Reference Detector”, 
the CEPC study group, arXiv:2510.05260

\bibitem{FCC-ee}
“FCC-ee: The Lepton Collider”, 
A. Abada et al., Eur. Phys. J. Spec. Top. 228, 261 (2019) 

\bibitem{BHLUMI404}
“Upgrade of the Monte Carlo program BHLUMI for Bhabha scattering 
at low angles to version 4.04”,
S. Jadash et al., Comp. Phys. Comm. Nucl. Instrum. 107, 229 (1997)

\bibitem{BHWIDE100}
“BHWIDE 1.00: O(alpha) YFS exponentiated Monte Carlo for Bhabha scattering 
at wide angles for LEP-1/SLC and LEP-2”,   
S. Jadach et al., Phys. Lett. B, 390, 298 (1997)

\bibitem{GEANT3}
R. Brun et al.,
``GEANT Detector Description and Simulation Tool'',
CERN-W5013, (1994).

\bibitem{GEANT4}
S. Agostinelli et al.",
``Geant4—a simulation toolkit'',
Nucl. Instrum. Methods A 506 (2003) 250.

\bibitem{YFS}
“The infrared divergence phenomena and high-energy processes”, 
D.R. Yennie et al., Annals Phys. 13, 379 (1961)


\end{thebibliography}



\end{document}